\documentclass[12pt,a4paper]{article}
\usepackage{a4wide,cite}
\newcommand{\be}{\begin{equation}}
\newcommand{\ee}{\end{equation}}
\newcommand{\bea}{\begin{eqnarray}}
\newcommand{\eea}{\end{eqnarray}}

\newcommand{\ep}{\varepsilon}
\newcommand{\nn}{\nonumber}

\newcommand{\Li}[2]{{\mbox{Li}}_{#1}\left(#2\right)}
\newcommand{\Cl}[2]{{\mbox{Cl}}_{#1}\left(#2\right)}

\catcode`\@=11
\begin{document}

 \thispagestyle{empty}
 \begin{flushright}
 {MZ-TH/98-48} \\[3mm]
 {hep-ph/9903328} \\[3mm]
 {March 1999}
 \end{flushright}
 \vspace*{2.0cm}
 \begin{center}
 {\Large \bf
 Threshold expansion of the sunset diagram}
 \end{center}
 \vspace{1cm}
 \begin{center}
 A.I.~Davydychev$^{a,b,}$\footnote{davyd@thep.physik.uni-mainz.de}
 \ \ and \ \
 V.A.~Smirnov$^{b,}$\footnote{smirnov@theory.npi.msu.su}
\\
 \vspace{1cm}
$^{a}${\em
 Department of Physics,
 University of Mainz, \\
 Staudingerweg 7,
 D-55099 Mainz, Germany}
\\
\vspace{.3cm}
$^{b}${\em
 Institute for Nuclear Physics,
 Moscow State University, \\
 119899, Moscow, Russia}
\end{center}
 \hspace{3in}
 \begin{abstract}
By use of the threshold expansion we develop an algorithm for analytical
evaluation, within dimensional regularization, of arbitrary terms in
the expansion of the (two-loop) sunset diagram with general masses 
$m_1$, $m_2$ and $m_3$ near its threshold,
i.e.\ in any given order in the difference between the external momentum
squared and its threshold value, $(m_1+m_2+m_3)^2$.
In particular, this algorithm includes an explicit recurrence procedure
to analytically calculate sunset diagrams with arbitrary
integer powers of propagators at the threshold.
 \end{abstract}

\newpage


{\large\bf \noindent 1. Introduction}

\vspace{3mm}

The purpose of our paper is to apply explicit prescriptions \cite{BS}
recently obtained for the expansion near threshold,
i.e.\ in powers of the difference between the external momentum squared
and its threshold value, to calculation of the sunset diagram
(see Fig.~1) with general masses and powers of propagators.
We thereby extend, to the threshold expansion,
analysis of two-loop Feynman integrals performed
in a series of papers \cite{TwoLoop1,BDS,TwoLoop2,TwoLoopExpOn} and
based on explicit formulae for the asymptotic expansion of Feynman
diagrams in various off-shell limits of momenta and masses
\cite{PrescrOff} 
(see also earlier papers \cite{PiTk} and operator analogues 
in~\cite{GLT}; for brief reviews, see in~\cite{BrRev}),
as well as for some typically Minkowskian on-shell
limits~\cite{PrescrOn}.
In particular, in ref.~\cite{BDS} the threshold behaviour
at small (as compared to some masses not involved in the cut)
non-zero thresholds was examined within the large mass expansion,
and the description of
three-particle thresholds depended crucially on a possibility
to describe the threshold behaviour of the sunset diagram
and a similar diagram with four propagators.

The sunset diagram (Fig.~1) represents the simplest example of
a diagram involving a three-particle threshold,
at $k^2=(m_1+m_2+m_3)^2$.
When one or two internal particles are massless, the
(four-dimensional) results can be obtained in terms of
dilogarithms \cite{B+ST,B+T}\footnote{Some other results for
two-loop two-point diagrams with masses can be found in
refs.~\cite{others}.}.
The situation gets more complicated when all three virtual particles
involved in the cut are massive. Such a situation occurs e.g.\
in the two-loop off-shell contributions to the Higgs self energy.
Although in three dimensions the result for this diagram
(even with different masses) is quite simple
\cite{Rajantie} (cf.\ also in ref.~\cite{GLP}),
in four dimensions no exact results in terms of known
functions (like polylogarithms, etc.) are available
in this (totally massive) case. Moreover, there are
arguments \cite{Scharf} that the result cannot be expressed
in terms of polylogarithms, except for the special cases
like the threshold (and pseudothresholds).

One possibility is to use various integral representations
\cite{num,B+T,Bau,PT}, in order to get numerical values for given 
masses and the momentum. Another possibility is to study analytic 
expansions in different regions. For instance,
using the algorithms presented in \cite{TwoLoop1}, one can construct
several terms of the small momentum expansion and the large
momentum expansion. Moreover, in ref.~\cite{BBBS} a closed
formula for the corresponding coefficients (in the case of
the sunset diagram) was derived, and the occurring
hypergeometric series were identified as Lauricella functions
of three variables. Nevertheless,
because of difficulties in constructing analytic continuation
of these functions, one cannot extract much information
about the threshold behaviour.
We also note that in ref.~\cite{Rem} a differential equation for
the sunset diagram was constructed, whose derivation was
essentially based on the integration by parts \cite{ibp} and
a recurrence procedure described in \cite{Tarasov}.
The first step towards explicit evaluation of the coefficients
of the threshold expansion was done in ref.~\cite{BDU}, where the
threshold (and pseudothreshold) values of the sunset diagram
(with unit powers of
the propagators, and also in the case when one of the powers
is equal to two) were calculated analytically in terms of dilogarithms
of the mass ratios.


The sunset diagram with general powers of propagators and masses
is given by the following two-loop integral (cf.\ Fig.~1):
\be
\label{def_L}
L(n; \nu_1,\nu_2,\nu_3) \equiv
\int\int \frac{\mbox{d}^n p_1 \; \mbox{d}^n p_2}
              {\left[ p_1^2 - m_1^2 \right]^{\nu_1}
               \left[ p_2^2 - m_2^2 \right]^{\nu_2}
               \left[ (k-p_1-p_2)^2-m_3^2 \right]^{\nu_3}} \; ,
\ee
where $n=4-2\ep$ is the space-time dimension, in the framework
of dimensional regularization \cite{dimreg}.
We shall use the notation
\be
\label{xi_i}
\mu_T\equiv m_1+m_2+m_3 \ , \hspace{8mm} 
\xi_i \equiv \frac{m_i}{\mu_T},
\ee
with $\sum_{i=1}^3 \xi_i = 1$, and
introduce the expansion parameter
\be
\label{def_y}
y\equiv \mu_T^2-k^2 \ .
\ee
To make the dependence on $y$ manifest, it is convenient to shift the
loop momenta in eq.~(\ref{def_L}) as
$p_1\to p_1+\xi_1 k, \; p_2\to p_2+\xi_2 k$.
Thus the r.h.s.\ of eq.~(\ref{def_L}) becomes
\be
\int \int \frac{\mbox{d}^n p_1 \; \mbox{d}^n p_2}
{\left[ p_1^2+2\xi_1 (k p_1) - \xi_1^2 y\right]^{\nu_1}
\left[ p_2^2+2\xi_2 (k p_2) - \xi_2^2 y\right]^{\nu_2}
\left[ (p_1+p_2)^2-2\xi_3 (k,p_1+p_2)-\xi_3^2 y\right]^{\nu_3}} \, .
\label{SS1}
\ee

Without loss of generality, in the time-like region we can choose
a frame with the external momentum $k=(k_0,\vec{0})$ with $k_0>0$.
According to the general prescriptions of the threshold expansion 
\cite{BS}, one should consider every loop momentum $p_i$ ($i=1,2$)
to be of the following four types:
\begin{eqnarray}
\mbox{{\em hard} (h):} && p_{i0}\sim |k|,\,\,|\vec{p}_i|\sim |k|,
\nn \\
\mbox{{\em soft} (s):} && p_{i0}\sim \sqrt{y},\,\,
|\vec{p}_i|\sim\sqrt{y},
\nn \\
\mbox{{\em potential} (p):} && p_{i0}\sim y/|k|,\,\,
|\vec{p}_i|\sim\sqrt{y},
\nn \\
\mbox{{\em ultrasoft} (us):} && p_{i0}\sim y/|k|,\,\,
|\vec{p}_i|\sim y/|k| \, ,
\nonumber
\end{eqnarray}
where $|k|\equiv\sqrt{k^2}=k_0$.

We shall consider the most complicated case when
all three masses are non-zero. Then only (h-h) and (p-p) regions
contribute to the threshold expansion because, for any other
region, one obtains scaleless integrals which are naturally put to zero
(they are analogous to massless tadpoles in dimensional 
regularization \cite{dimreg}).
Note that if one or two masses were zero then we would have non-zero
(us-p) or (us-us) contributions, instead of the (p-p) one
(cf.\ in refs.~\cite{BS,CM+BSS}).

As the main example, we shall treat the diagram with
$\nu_1=\nu_2=\nu_3=1$.
Within threshold expansion, this ``master'' sunset diagram is
represented as
\be
\label{hhpp}
L(4-2\ep; 1,1,1)
= -\pi^{4-2\ep}
\; \Gamma(1+2\ep) \;  k^2
\sum_{j=0}^{\infty}
\left( \frac{y}{k^2}\right)^j
\left[ C_j^{{\rm (h-h)}} (k^2)^{-2\ep}+
C_j^{{\rm (p-p)}} y^{-2\ep} \right] \; ,
\ee
where we have extracted the factor $\mbox{i}\pi^{n/2}$ per loop.
Our results can be easily generalized for sunset diagrams
with any indices $\nu_i$.

The (h-h) contribution is given by expanding the integrand in (\ref{SS1})
in Taylor series in $y$, so that these are sunset integrals,
with various (higher) indices of the lines, evaluated exactly
at the threshold ($y=0$). In sections~2--4
we shall describe how an arbitrary integral of this
kind can be analytically evaluated.
It happens that general solutions of recurrence relations
constructed in \cite{Tarasov} cannot be directly applied
at the threshold. Nevertheless, those relations are useful
at the threshold, providing equations which should be used
together with the usual integration-by-parts relations \cite{ibp}.

The (p-p) contribution is given by expanding the propagators
in (\ref{SS1}) in Taylor series in the squares of time components,
$p_{10}^2$, $p_{20}^2$ and $(p_{10}+p_{20})^2$.
The evaluation of the (p-p) part of the expansion is presented
in Section~5. 
In Section~6 we compare our results for the threshold expansion 
with results based on numerical integration.
We briefly summarize our results in Section~7.

In general, the threshold expansion has a complicated structure
and it is difficult to see how it works.
Nevertheless, in our particular example we can present
some simple arguments. Namely, it happens that the remainder
of the expansion of the sunset diagram can be constructed by 
applying, to eq.~(\ref{SS1}), an operator
\be
{\cal R}^{(N)}=\left(1-{\cal T}_y^{(N)}\right)
\left(1-{\cal T}_{p_{i0}^2}^{(N-2)} \right) .
\ee
Here ${\cal T}_{y}^{(N)}$ denotes an operator that
picks up the first $N$ terms of Taylor expansion in $y$
(cf.\ eq.~(\ref{exp1}) below),
whereas ${\cal T}_{p_{i0}^2}^{(N)}$ corresponds to the
first $N$ terms of the expansion in squares of time
components $p_{10}^2$, $p_{20}^2$ and $(p_{10}+p_{20})^2$
(cf.\ eq.~(\ref{exp2}) below).
It is assumed that these operators ``commute'' with the
the loop integrations, i.e.\ they should be applied directly
to the denominators in eq.~(\ref{SS1}).

For instance, for the integral with $\nu_i=1$ we can write
\bea
\label{rem}
L(n;1,1,1) &=& \left(1-{\cal R}^{(N)}\right) L(n;1,1,1) 
+ {\cal R}^{(N)} L(n;1,1,1) \,
\nn \\
 &=& \left({\cal T}_y^{(N)}+{\cal T}^{(N-2)}_{p_{i0}^2} 
           - {\cal T}^{(N)}_y {\cal T}^{(N-2)}_{p_{i0}^2}
\right) L(n;1,1,1) 
+ {\cal R}^{(N)} L(n;1,1,1) \, ,
\hspace{5mm}
\eea
where the terms ${\cal T}_y^{(N)}L$ and ${\cal T}^{(N-2)}_{p_{i0}^2}L$ 
give us the (h-h) and (p-p) contributions
to the threshold expansion (\ref{hhpp}), respectively.
For dimensionally-regularized integrals, the product of these operators,
${\cal T}^{(N)}_y {\cal T}^{(N-2)}_{p_{i0}^2}L$, vanishes. 
To see this, one can integrate in $p_{10}$ and $p_{20}$ 
using residual theorem:
then, the resulting integrals in $\vec{p}_1$ and $\vec{p}_2$ 
are scaleless (i.e.\ analogous to massless tadpoles, but 
in $(n-1)$ dimensions), and therefore vanish.
 
Finally, the last term on the r.h.s.\ of eq.~(\ref{rem}),
${\cal R}^{(N)} L(n;1,1,1)$, plays the role 
of the remainder of the expansion. To show this, consider first
the region of small loop momenta where the operator ${\cal T}^{(N)}_y$ 
is dangerous because it produces infrared threshold singularities. 
However these singularities are removed by the operator 
$\left(1-{\cal T}_{p_{i0}^2}^{(N-2)}\right)$ which increases
powers of the variables $p_1$ and $p_2$ in the numerator.
  Then, consider the region of large loop momenta where the operator
  ${\cal T}^{(N-2)}_{p_{i0}^2}$ is dangerous because it
  generates specific divergences (corresponding to ultraviolet 
  singularities in lower dimensions). Still, the latter
  are removed by the operator $\left(1-{\cal T}_y^{(N)}\right)$
  which effectively increases the powers of integration momenta
  in the denominators.
Thus the remainder ${\cal R}^{(N)}$ does not involve new divergences,
neither ultraviolet nor infrared ones, as compared to
the initial Feynman integral (\ref{SS1}). 
Moreover, taking into account the operator
$\left(1-{\cal T}_y^{(N)}\right)$ we see that the order 
of this remainder is $y^{N+1}$, up to logarithms.

\vspace{5mm}

{\large\bf \noindent 2. Taylor expansion and the (h-h) contribution}

\vspace{3mm}

The formal expansion in $y$ of the denominators in the integrand
of eq.~(\ref{SS1}) can be performed via
\be
\label{exp1}
\frac{1}{\left[p_i^2+2\xi_i(kp_i)-\xi_i^2 y\right]^{\nu_i}}
=\sum_{j_i=0}^{\infty}\frac{(\nu_i)_{j_i}}{j_i!}\;
\frac{\xi_i^{2j_i}\; y^{j_i}}
{\left[p_i^2+2\xi_i(kp_i)\right]^{\nu_i+j_i}} \; ,
\ee
where $(\nu)_j\equiv\Gamma(\nu+j)/\Gamma(\nu)$ is the Pochhammer
symbol. If we denote $p_3=-p_1-p_2$, the
third denominator can also be taken into account by eq.~(\ref{exp1}).
Then, collecting the terms with given powers of $y$,
we get the (h-h) contributions (cf.\ eq.~(\ref{hhpp})).
According to the r.h.s.\ of eq.~(\ref{exp1}), each separate term
corresponds to the integral (\ref{SS1}) with shifted powers
of propagators ($\nu_i\to\nu_i+j_i$) and $y=0$, i.e.\
at the threshold.

One should however keep in mind
which variables are considered as independent ones.
Let us consider $\xi_i$, eq.~(\ref{xi_i}), as ``external'' parameters
characterizing the mass ratios. Then
the remaining variables are $y$, $k^2$ and $\mu_T^2=(m_1+m_2+m_3)^2$,
which are dependent, according to the definition (\ref{def_y}).
Since $y$ is the main parameter of the threshold expansion
(cf.\ eq.~(\ref{hhpp})), we can choose between
the following pairs of independent variables: (i) $y$ and $k^2$,
and (ii) $y$ and $\mu_T^2$.

On one hand, in the context of the threshold expansion
\cite{BS} (including the related issue of the non-relativistic limit),
the first set, $y$ and $k^2$, looks more natural (cf.\ eq.~(\ref{hhpp})).
In particular, the momentum integrals arising after expansion
in $y$ (\ref{exp1}) are functions of $k^2$.
In section~5, we shall also see that such a choice is more
convenient for evaluating the (p-p) contribution.
On the other hand, when evaluating the threshold values of
the momentum-space integrals (corresponding to the (h-h)
contribution), it is convenient
to put $k^2=\mu_T^2$, i.e.\ to calculate
\be
\label{L_T}
L_T(n; \nu_1,\nu_2,\nu_3)\equiv
\left. L(n; \nu_1,\nu_2,\nu_3)\right|_{k^2=(m_1+m_2+m_3)^2} .
\ee
This would correspond to the second set of independent variables,
$y$ and $\mu_T^2$.
In particular, when $\nu_i=1$ we define
\be
\label{hhpp2}
L(4-2\ep; 1,1,1)
= -\pi^{4-2\ep} \Gamma^2(1+\ep)\;
\mu_T^2
\sum_{j=0}^{\infty}
\left( \frac{y}{\mu_T^2}\right)^j
\left[ {\widetilde{C}}_j^{{\rm (h-h)}}
\left(\mu_T^2\right)^{-2\ep}+
{\widetilde{C}}_j^{{\rm (p-p)}} y^{-2\ep} \right] \; .
\ee
In such way, it is
more straightforward to use the threshold results for the
lowest integrals obtained in \cite{BDU}. Another advantage,
as we shall see, is connected with the ``book keeping'' of
ultraviolet singularities.

Anyway, the transition between these two options
(and, in particular, between the coefficients $C_j$
and ${\widetilde{C}}_j$ in eqs.~(\ref{hhpp}) and (\ref{hhpp2}),
respectively)
is rather straightforward. For instance, any integral $L_T$
(\ref{L_T}) can be presented as
$\left(\mu_T^2\right)^{n-\nu_1-\nu_2-\nu_3}$
times a function of dimensionless parameters $\xi_i$,
eq.~(\ref{xi_i}).
When we consider the momentum integrals corresponding
to the expansion (\ref{exp1}), the only difference is that
we need to replace $\mu_T^2$ by $k^2$, namely:
\bea
\label{transition}
\int\int \frac{{\mbox{d}}^n p_1 \;{\mbox{d}}^n p_2}
{\left[ p_1^2+2\xi_1(kp_1) \right]^{\nu_1}
 \left[ p_2^2+2\xi_2(kp_2) \right]^{\nu_2}
 \left[ (p_1+p_2)^2-2\xi_3(k,p_1+p_2) \right]^{\nu_3}}
\nn \\
=( k^2/\mu_T^2)^{n-\nu_1-\nu_2-\nu_3}
\; L_T(n; \nu_1,\nu_2,\nu_3) \; .
\hspace{5mm}
\eea

If we consider the (h-h) contributions (\ref{exp1}) corresponding to
$L(n;1,1,1)$, the first terms are proportional to
\bea
\label{hhterms}
&& L_T(n;1,1,1),
\nn \\
&& m_1^2 L_T(n;2,1,1) +m_2^2 L_T(n;1,2,1) + m_3^2 L(n;1,1,2),
\nn \\
&& m_1^2 m_2^2 L_T(n;2,2,1)
+ m_2^2 m_3^2 L_T(n;1,2,2)
+ m_3^2 m_1^2 L_T(n;2,1,2)
\nn \\
&& \hspace*{10mm}
+ m_1^4 L_T(n;3,1,1) + m_2^4 L_T(n;1,3,1)
+ m_3^4 L_T(n;1,1,3),
\eea
etc. It is easy to see that each new order will contain
ultraviolet-divergent combinations
\be
m_1^{2j} L_T(n;j+1,1,1) + m_2^{2j} L_T(n;1,j+1,1)
+ m_3^{2j} L_T(n;1,1,j+1) .
\ee

The situation is different if we use the variables
$y$ and $\mu_T^2$, according to eq.~(\ref{hhpp2}).
It is instructive to consider
the formal Taylor expansion around the threshold
$k^2=\mu_T^2$ ($y=0$),
\be
\label{Tay0}
\sum_{j=0}^{\infty}
\frac{(-y)^j}{j!}
\left[
\left(\frac{\partial}{\partial k^2}\right)^j
L(n;\nu_1,\nu_2,\nu_3)\right]_{k^2=(m_1+m_2+m_3)^2},
\ee
which also corresponds to the (h-h) contribution.
An algorithmically convenient way to construct expressions
for the derivatives in (\ref{Tay0}) is to use the formulae
\be
\label{Tay}
\left(\frac{\partial}{\partial k^2}\right)^j
L(n;\nu_1,\nu_2,\nu_3)
=\frac{(-1)^j}{\pi^{2j}}\;
(\nu_1)_j (\nu_2)_j (\nu_3)_j \;
L(n+2j;\nu_1+j,\nu_2+j,\nu_3+j)
\ee
and
\bea
\label{n->n-2}
\pi^{-2} k^2 (\nu_1+j-1) (\nu_2+j-1) (\nu_3+j-1)
L(n+2j;\nu_1+j,\nu_2+j,\nu_3+j)
\hspace{20mm}
\nn \\
= (\nu_1+\nu_2+\nu_3-n+j-1)\;
L(n+2j-2;\nu_1+j-1,\nu_2+j-1,\nu_3+j-1)
\nn \\
+(\nu_1+j-1) m_1^2\; L(n+2j-2;\nu_1+j,\nu_2+j-1,\nu_3+j-1)
\hspace{6mm}
\nn \\
+(\nu_2+j-1) m_2^2\; L(n+2j-2;\nu_1+j-1,\nu_2+j,\nu_3+j-1)
\hspace{6mm}
\nn \\
+(\nu_3+j-1) m_3^2\; L(n+2j-2;\nu_1+j-1,\nu_2+j-1,\nu_3+j) .
\hspace{4mm}
\eea
Applying eq.~(\ref{n->n-2}) $j$ times, we reduce the space-time dimension
back to $n=4-2\ep$.
Both formulae (\ref{Tay}) and (\ref{n->n-2}) can be easily
derived by using the modified Feynman parametric
representation for the integral (\ref{def_L})
given by eq.~(4) of ref.~\cite{BDU} (cf.\ also in
refs.~\cite{Tarasov2}).

When we consider the Taylor expansion of $L(n;1,1,1)$,
the derivatives are given by
\be
\label{Tay1}
-k^2\frac{\partial}{\partial k^2} L(n; 1,1,1)
= (3\!-\!n) L(n;1,1,1) + m_1^2 L(n;2,1,1)
+m_2^2 L(n;1,2,1) + m_3^2 L(n;1,1,2),
\ee
\bea
\label{Tay2}
{\textstyle{1\over2}} 
(k^2)^2 \left( \frac{\partial}{\partial k^2} \right)^2
L(n; 1,1,1)
&=& {\textstyle{1\over2}} (4-n)(3-n) L(n;1,1,1)
\nn \\
&& +(4\!-\!n)
\left[ m_1^2 L(n;2,1,1)\!+ m_2^2 L(n;1,2,1)\!+ m_3^2 L(n;1,1,2)
\right]
\nn \\
&& + m_1^2 m_2^2 L(n;2,2,1)\! + m_2^2 m_3^2 L(n;1,2,2)\!
+m_3^2 m_1^2 L(n;2,1,2)
\nn \\
&& + m_1^4 L(n;3,1,1) + m_2^4 L(n;1,3,1) + m_3^4 L(n;1,1,3),
\eea
etc. In these derivatives (see eq.~(\ref{Tay0})), we should put
$k^2=(m_1+m_2+m_3)^2$ ($L\to L_T$).

Note that now the ultraviolet singularities are present only in
the $y^0$ and $y^1$ terms of the expansion of $L(n;1,1,1)$.
One can see that, apart from the lower integrals, the
derivatives (\ref{Tay1}) and (\ref{Tay2}) contain the same
combinations as (\ref{hhterms}). Furthermore, one can get
the results (\ref{Tay1}) and (\ref{Tay2}) directly from
eq.~(\ref{hhterms}), by multiplying each contribution
by the factor (\ref{transition}),
$(k^2/\mu_T^2)^{n-3}=\left(1-y/\mu_T^2\right)^{n-3}$,
expanding it in $y$, and collecting terms at given powers of $y$.

The analytic results for $L_T(4-2\ep;1,1,1)$ and
$L_T(4-2\ep;1,1,2)$, expanded in $\ep$ up to the finite part,
are given in eqs.~(23) and (22) of ref.~\cite{BDU},
respectively. The results for
$L_T(4-2\ep;2,1,1)$ and $L_T(4-2\ep;1,2,1)$ can be obtained
from $L_T(4-2\ep;1,1,2)$ via permutation of the masses $m_i$.
Those results involve the following functions:
\bea
\label{T-}
T^{-}(z) &=& \Li{2}{-z} - \Li{2}{-1/z}
       +2\ln z \ln(1+z) - \ln^2 z
\nn \\
&=& 2 \Li{2}{-z} + {\textstyle{1\over6}}\pi^2 + 2\ln z \ln(1+z)
    - {\textstyle{1\over2}} \ln^2 z
\nn \\
&=& 2 \Li{2}{ 1/(1+z) } -  {\textstyle{1\over6}}\pi^2
    + \ln^2(1+z) - {\textstyle{1\over2}} \ln^2 z ,
\eea
\be
\label{theta_i}
\theta_i \equiv
\arctan\left( m_i \sqrt{\frac{m_1+m_2+m_3}{m_1 m_2 m_3}} \right) ,
\hspace{10mm}
\theta_1+\theta_2+\theta_3=\pi .
\ee
Note that the arguments of $T^{-}$ are just the mass
ratios $m_j/m_l$, and the following inversion property
is valid: $T^{-}(m_j/m_l)=-T^{-}(m_l/m_j)$.
In particular, $T^{-}(1)=0$ (remember that 
$\Li{2}{-1}=-{\textstyle{1\over12}}\pi^2$). 
Similar functions have also appeared in refs.~\cite{dilogs}.

In the next two sections, we shall consider
evaluation of the integrals $L_T$ with higher powers of
propagators.

\newpage

{\large\bf \noindent 3. Recurrence relations at the threshold}

\vspace{3mm}

In this section, we shall consider relations connecting
the integrals (\ref{def_L}) with different indices $\nu_i$.
When $L$ is used without its arguments, the
non-shifted $n$ and $\nu_i$ are understood.
The standard notation for the raising and lowering operators reads
\be
\label{123pm}
{\mbox{\bf 1}}^{\pm}L
= L(n; \nu_1\pm 1,\nu_2,\nu_3), \hspace{5mm}
{\mbox{\bf 2}}^{\pm}L
= L(n; \nu_1,\nu_2\pm 1,\nu_3), \hspace{5mm}
{\mbox{\bf 3}}^{\pm}L
= L(n; \nu_1,\nu_2,\nu_3\pm 1).
\ee
When the external momentum squared takes its threshold value,
$k^2=(m_1+m_2+m_3)^2$, we shall use the notation (\ref{L_T}).

Using the identities \cite{ibp}
\be
\label{ibp_12}
\int\int \mbox{d}^n p_1 \; \mbox{d}^n p_2
\frac{\partial}{\partial {p_i}_{\mu}}
\left\{
\frac{A_{\mu}^{(j)}}
{\left[p_1^2-m_1^2\right]^{\nu_1}
\left[p_2^2-m_2^2\right]^{\nu_2}
\left[(k-p_1-p_2)^2-m_3^2\right]^{\nu_3}}
\right\} = 0,
\ee
with $A_{\mu}^{(j)}=\left\{{p_1}_{\mu},{p_2}_{\mu},{k}_{\mu}\right\}$,
we obtain a set of integration-by-parts relations.
For example, one of the equations can be presented as
\bea
\label{ibp1pp}
\left[ 2m_1^2\nu_1{\mbox{\bf 1}}^{+}
+(m_1^2+m_2^2+m_3^2-k^2)\nu_2{\mbox{\bf 2}}^{+}
+ 2m_3^2\nu_3{\mbox{\bf 3}}^{+}
+\frac{2m_1^2\nu_1\nu_2}{\nu_3-1}\;
{\mbox{\bf 3}}^{-}{\mbox{\bf 1}}^{+}{\mbox{\bf 2}}^{+}
\right] L
\nn \\
=\left[ 2n-2\nu_1-\nu_2-2\nu_3
-\nu_2{\mbox{\bf 1}}^{-}{\mbox{\bf 2}}^{+}
+\frac{n-2\nu_1-\nu_3+1}{\nu_3-1}\nu_2
{\mbox{\bf 3}}^{-}{\mbox{\bf 2}}^{+}
\right] L ,
\eea
and five other equations can be obtained from (\ref{ibp1pp})
by permutations of the indices (1,2,3).

Using three of these six equations, we can e.g.\ express
${\mbox{\bf 3}}^{-}{\mbox{\bf 1}}^{+}{\mbox{\bf 2}}^{+}L$,
${\mbox{\bf 2}}^{-}{\mbox{\bf 1}}^{+}{\mbox{\bf 3}}^{+}L$ and
${\mbox{\bf 1}}^{-}{\mbox{\bf 2}}^{+}{\mbox{\bf 3}}^{+}L$
in terms of ${\mbox{\bf 1}}^{+}L$,
${\mbox{\bf 2}}^{+}L$ and ${\mbox{\bf 3}}^{+}L$.
However, three remaining equations
for  ${\mbox{\bf 1}}^{+}L$,
${\mbox{\bf 2}}^{+}L$ and ${\mbox{\bf 3}}^{+}L$
happen to be linearly dependent, for arbitrary $k^2$.
Useful corollaries of eq.~(\ref{ibp1pp}) and its permutations
are
\be
\label{++13}
\left[
m_1^2\nu_1(\nu_1\!+\!1) {\mbox{\bf 1}}^{+}{\mbox{\bf 1}}^{+}
- m_3^2\nu_3(\nu_3\!+\!1) {\mbox{\bf 3}}^{+}{\mbox{\bf 3}}^{+}
\right] L
={\textstyle{1\over2}}\left[
(n\!-\!2\nu_1\!-\!2)\nu_1{\mbox{\bf 1}}^{+}
-(n\!-\!2\nu_3\!-\!2)\nu_3{\mbox{\bf 3}}^{+}
\right] L,
\ee
\be
\label{++23}
\left[
m_2^2\nu_2(\nu_2\!+\!1) {\mbox{\bf 2}}^{+}{\mbox{\bf 2}}^{+}
-m_3^2\nu_3(\nu_3\!+\!1) {\mbox{\bf 3}}^{+}{\mbox{\bf 3}}^{+}
\right] L
={\textstyle{1\over2}}\left[
(n\!-\!2\nu_2\!-\!2)\nu_2{\mbox{\bf 2}}^{+}
-(n\!-\!2\nu_3\!-\!2)\nu_3{\mbox{\bf 3}}^{+}
\right] L .
\ee
Below we shall use these two equations as a part of the main
algorithm.

Let us denote $\sigma\equiv\nu_1+\nu_2+\nu_3$.
Since the threshold values of the integrals
with $\sigma=3$ ($\nu_1=\nu_2=\nu_3=1$) and
$\sigma=4$ ($\nu_1=\nu_2=1$, $\nu_3=2$ and permutations)
are known \cite{BDU}, the first non-trivial step
is to get results for six integrals with $\sigma=5$
($L(n;2,2,1)$, $L(n;1,1,3)$ and permutations).
Introduce the following notation for the symmetric sums of these
integrals:
\be
\label{S221}
S_{221}\equiv L(n;1,2,2) + L(n;2,1,2) + L(n;2,2,1),
\ee
\be
\label{S113}
S_{113}\equiv m_1^2 L(n;3,1,1) + m_2^2 L(n;1,3,1)
+ m_3^2 L(n;1,1,3).
\ee
Then, using eqs.~(\ref{++13})--(\ref{++23}) we get
\be
\label{L113_rec}
L(n;1,1,3)=\frac{1}{3m_3^2}\left\{ S_{113}
+{\textstyle{1\over4}} (n-4) 
\left[ 2 L(n;1,1,2)-L(n;2,1,1)-L(n;1,2,1) \right]
\right\}
\ee
and analogous results for $L(n;3,1,1)$ and $L(n;1,3,1)$.
Furthermore, using eq.~(\ref{ibp1pp}) and its permutations,
we get
\bea
\label{L221_rec}
L(n;2,2,1) &=& \frac{1}{k^2-m_1^2-m_2^2+m_3^2}
\bigg\{ 2 m_3^2 S_{221} + {\textstyle{4\over3}} S_{113}
- {\textstyle{1\over3}} (n-4) L(n;1,1,2)
\nn \\
&& - {\textstyle{1\over3}} (4n-13) 
\left[ L(n;2,1,1) + L(n;1,2,1) \right]
 + L(n;2,2,0) \bigg\}
\hspace{5mm}
\eea
and analogous results for $L(n;1,2,2)$ and $L(n;2,1,2)$.
Note that at the threshold
\[
\left. \left( k^2-m_1^2-m_2^2+m_3^2
       \right)\right|_{k^2=(m_1+m_2+m_3)^2}
=2(m_1+m_3)(m_2+m_3) .
\]

Furthermore, taking the sum of eq.~(\ref{L221_rec}) and
its permutations,
we can express $S_{221}$ via $S_{113}$ and lower integrals.
In particular, at the threshold we get
\bea
\label{s1}
m_1 m_2 m_3 S_{221}^{(T)}
\!-\! {\textstyle{2\over3}} \mu_T S_{113}^{(T)}
= -{\textstyle{1\over12}}(5n\!-\!17) \mu_T
\left[ L_T(n;2,1,1)\!+\!L_T(n;1,2,1)\!+\!L_T(n;1,1,2) \right]
\nn \\
-{\textstyle{1\over4}}(n-3)
\left[ m_1 L_T(n;2,1,1)+m_2 L_T(n;1,2,1)
      + m_3 L_T(n;1,1,2) \right]
\hspace{20mm}
\nn \\
+ {\textstyle{1\over4}}
\left[ (m_2+m_3)L(n;0,2,2)+(m_1+m_3)L(n;2,0,2)+(m_1+m_2)L(n;2,2,0)
\right],
\hspace{5mm}
\eea
where the superscript ``$(T)$'' means that the corresponding
sums, $S^{(T)}_{221}$ and $S^{(T)}_{113}$,
are considered at the threshold.
Therefore, using the integration-by-parts relations we
reduce all the six integrals with $\sigma=5$
to a single unknown function (say, $S_{221}$).

A missing ``block'' completing the algorithm can be
obtained using some equations
presented in ref.~\cite{Tarasov}.
Consider eqs.~(74) and (81) of \cite{Tarasov}.
Their l.h.s.'s contain a factor
\bea
\label{D123}
D_{123} &\equiv& \left[ k^2 - (m_1+m_2+m_3)^2 \right]
\left[ k^2 - (-m_1+m_2+m_3)^2 \right]
\nn \\
&& \times
\left[ k^2 - (m_1-m_2+m_3)^2 \right]
\left[ k^2 - (m_1+m_2-m_3)^2 \right] .
\eea
In the case of interest (i.e.\ at the threshold) these l.h.s.'s
vanish, since $D_{123}=0$.
Therefore, at the threshold these equations
cannot be used in the way as it is suggested in ref.~\cite{Tarasov},
since this is a degenerate case.
Nevertheless, the fact that the r.h.s.'s should also be
equal to zero, provides some non-trivial conditions
on the integrals involved. It is interesting that at the threshold
both eqs.~(74) and (81) of \cite{Tarasov} lead to the same condition:
\bea
\label{Tar}
\left\{
\nu_1 m_1 \left[ (2n-\nu_1-2\nu_2-2\nu_3-1)m_1
+(n-\nu_2-2\nu_3)m_2+(n-2\nu_2-\nu_3)m_3 \right]
                         {\mbox{\bf 1}}^{+}
\right.
\nn \\
+\nu_2 m_2 \left[ (n-\nu_1-2\nu_3)m_1+(2n-2\nu_1-\nu_2-2\nu_3-1)m_2
                 +(n-2\nu_1-\nu_3)m_3 \right]
                         {\mbox{\bf 2}}^{+}
\nn \\
+\nu_3 m_3 \left[ (n-\nu_1-2\nu_2)m_1+(n-2\nu_1-\nu_2)m_2
                 +(2n-2\nu_1-2\nu_2-\nu_3-1)m_3 \right]
                         {\mbox{\bf 3}}^{+}
\nn \\
-\nu_2 \nu_3 m_2 m_3
{\mbox{\bf 1}}^{-}{\mbox{\bf 2}}^{+}{\mbox{\bf 3}}^{+}
-\nu_1 \nu_3 m_1 m_3
{\mbox{\bf 2}}^{-}{\mbox{\bf 1}}^{+}{\mbox{\bf 3}}^{+}
-\nu_1 \nu_2 m_1 m_2
{\mbox{\bf 3}}^{-}{\mbox{\bf 1}}^{+}{\mbox{\bf 2}}^{+}
\hspace{20mm}
\nn \\
\left.
-{\textstyle{1\over2}}(n-\nu_1-\nu_2-\nu_3)
(3n-2\nu_1-2\nu_2-2\nu_3-2)
\right\} L_T(n;\nu_1,\nu_2,\nu_3) = 0 .
\eea
For example, when $\nu_1=\nu_2=\nu_3=1$, eq.~(\ref{Tar}) yields
\bea
\label{Tar111}
(n-3)\left\{ m_1(2m_1+m_2+m_3)L_T(n;2,1,1)
           + m_2(m_1+2m_2+m_3)L_T(n;1,2,1)
\right.
\nn \\
\left.
           + m_3(m_1+m_2+2m_3)L_T(n;1,1,2)
-{\textstyle{1\over2}} (3n-8) L_T(n;1,1,1) \right\}
\nn \\
=m_2 m_3 L(n;0,2,2) + m_1 m_3 L(n;2,0,2) + m_1 m_2 L(n;2,2,0),
\eea
where on the r.h.s.\ we have just the products of massive tadpoles,
\be
\label{tadpoles}
L(n;\nu_1,\nu_2,0)=
\mbox{i}^{2-2\nu_1-2\nu_2} \pi^n \;
(m_1^2)^{n/2-\nu_1} \; (m_2^2)^{n/2-\nu_2} \;
\frac{\Gamma(\nu_1-n/2) \Gamma(\nu_2-n/2)}
{\Gamma(\nu_1) \Gamma(\nu_2)}
\ee
and permutations.
We have checked that the results
for $L_T(n;1,1,1)$, $L_T(n;1,1,2)$ (and permutations)
presented in eqs.~(22) and (21) of \cite{BDU}
satisfy eq.~(\ref{Tar111}).

Furthermore, considering eq.~(\ref{Tar}) for $\nu_1=\nu_2=1$,
$\nu_3=2$ (and permutations), we obtain an extra 
(independent of (\ref{s1})) condition
on the sums $S_{221}^{(T)}$ and $S_{113}^{(T)}$.
In this way we can obtain results for the integrals
$L_T(n;2,2,1)$, $L_T(n;1,1,3)$ (and permutations)
in terms of lower integrals. However, at this step
(solving the system of linear equations)
we get the factor
$(n-4)$ in the denominator. This means that we need to
know the $\ep$ part of the lower integrals, in order to
get the finite part of $L_T(n;2,2,1)$, etc.
Alternatively, we can calculate the finite part of
$L_T(n;2,2,1)$ via straightforward calculation.
As we shall see, this will be enough to calculate the
higher integrals (see in Appendix~A for details).

\vspace{5mm}

{\large\bf \noindent 4. Calculation of $L_T(4-2\ep;2,2,1)$ 
                        with different masses}

\vspace{3mm}

We start from the two-fold integral representation (5) of
ref.~\cite{BDU}, which in the case $\nu_1=\nu_2=2$, $\nu_3=1$
reads
\bea
\label{ChWu}
L_T(4-2\ep;2,2,1)=\pi^{4-2\ep}\;\Gamma(1+2\ep)\; (m_1 m_2)^{-\ep} \;
m_3^{1-\ep}
\hspace{58mm}
\nn \\
\times\! \int\limits_0^{\infty}\! \int\limits_0^{\infty}
\frac{\mbox{d}\xi \; \mbox{d}\eta \; \xi^{\ep} \; \eta^{\ep}}
{(m_1 \xi + m_2 \eta + m_3)^{1-3\ep}
\left[ m_2 m_3 \xi (1\!-\!\eta)^2 + m_1 m_3 \eta (1\!-\!\xi)^2
     + m_1 m_2 (\xi \!-\!\eta)^2 \right]^{1+2\ep}} .
\hspace{4mm}
\eea
The threshold singularity originates from the region $\xi\sim\eta\sim 1$.
Indeed, analyzing eq.~(5) of \cite{BDU} we see that the threshold
singularity takes place (in the four-dimensional space) when
$\nu_1+\nu_2+\nu_3\geq 5$. It is of an infrared origin and (as we shall
see) appears in dimensionally-regularized integrals as a pole in
$\ep=(4-n)/2$.

The double integral in the second line of
eq.~(\ref{ChWu}) can be decomposed as
\bea
\label{ChWu2}
\frac{1}{\mu_T^{1-3\ep}}
\int\limits_0^{\infty} \int\limits_0^{\infty}
\frac{\mbox{d}\xi \; \mbox{d}\eta}
{\left[ m_2 m_3 \xi (1-\eta)^2 + m_1 m_3 \eta (1-\xi)^2
     + m_1 m_2 (\xi -\eta)^2 \right]^{1+2\ep}}
\hspace{32mm}
\nn \\
+ \frac{1}{\mu_T}
\int\limits_0^{\infty} \int\limits_0^{\infty}
\frac{\mbox{d}\xi \; \mbox{d}\eta}
{\left[ m_2 m_3 \xi (1\!-\!\eta)^2 + m_1 m_3 \eta (1\!-\!\xi)^2
     + m_1 m_2 (\xi \!-\!\eta)^2 \right]}\;
\frac{\left[ m_1(1\!-\!\xi)+m_2(1\!-\!\eta)\right]}
{(m_1 \xi + m_2 \eta + m_3)}
\nn \\
+ {\cal{O}}(\ep) ,
\hspace{5mm}
\eea
where the threshold singularity is only in the first term
(which is simpler than the original integral (\ref{ChWu})),
whereas the second term is finite as $n\to4$.

To evaluate the integrals in (\ref{ChWu2}), we can proceed 
in the same way as in \cite{BDU}. 
However, the following substitution of variables\footnote{A.~D.
is grateful to J.B.~Tausk for discussion of this substitution
in connection with ref.~\cite{BDU}.}
happens to be more efficient than eq.~(13) in \cite{BDU}:
\be
\label{JBT}
\xi=1+(m_2/m_3)\lambda z, \hspace{10mm}
\eta=1+(m_1/m_3) \lambda (1-z) \; .
\ee
Then, the region of integration in variables $z, \lambda$
is defined by
\begin{eqnarray*}
-m_3/(m_2 \lambda) \leq & z & \leq 1 + m_3/(m_1 \lambda) 
\hspace{8mm} (\mbox{for}\;\;\; \lambda>0 ) \ ,
\\
1 + m_3/(m_1 \lambda) \leq & z & \leq -m_3/(m_2 \lambda)
\hspace{12mm} (\mbox{for}\;\;\; \lambda_{\rm min}<\lambda<0 ) \ ,
\end{eqnarray*}
with $\lambda_{\rm min}=-m_3(m_1+m_2)/(m_1m_2)$.
Note that the resulting integrand is invariant under 
$(z,m_1)\leftrightarrow(1-z,m_2)$.
The threshold singularity corresponds to the region $\lambda\sim0$:
the transformed integrand of the first integral in 
(\ref{ChWu2}) contains $\lambda^{-1-4\ep}$. 

Evaluating the integrals in eq.~(\ref{ChWu2})
we arrive at the following result:
\bea
\label{Lt221}
L_T(4-2\ep;2,2,1)=
\pi^{4-2\ep}\;\Gamma^2(1+\ep)\;(m_1m_2m_3)^{-3\ep}\;
\mu_T^{-2+5\ep}
\hspace{49mm}
\nn \\
\times\left\{-m_3\sqrt{\frac{m_1+m_2+m_3}{m_1m_2m_3}}
\left[\frac{\pi}{2\ep}
+\pi {\cal{L}}(m_1,m_2,m_3)
+{\cal{C}}(\theta_1,\theta_2,\theta_3)
\right] \right.
\hspace{35mm}
\nn \\
\left. \left.
+T^{-}\left(\frac{m_1}{m_3}\right)
+T^{-}\left(\frac{m_2}{m_3}\right)
+\frac{4\pi^2}{3} \!-\! 4\pi\theta_3
+\ln^2\frac{m_1}{m_2} - {\textstyle{1\over2}}\ln^2\frac{m_1}{m_3}
-{\textstyle{1\over2}}\ln^2\frac{m_2}{m_3}
\right] \right\} + {\cal{O}}(\ep),
\hspace{4mm}
\eea
with 
\be
\label{Clausens}
{\cal{C}}(\theta_1,\theta_2,\theta_3)\equiv
\Cl{2}{2\theta_1}+\Cl{2}{2\theta_2}+\Cl{2}{2\theta_3}
+\Cl{2}{\pi\!-\!2\theta_1}
+\Cl{2}{\pi\!-\!2\theta_2}
+\Cl{2}{\pi\!-\!2\theta_3}
\ee
and
\be
\label{Logs}
{\cal{L}}(m_1,m_2,m_3)\equiv
\ln\left(\frac{(m_1+m_2)(m_2+m_3)(m_3+m_1)}{(m_1+m_2+m_3)^3}
\right) - 6\ln{2} .
\ee
The functions $T^{-}$ and $\theta_i$ are defined in eqs.~(\ref{T-})
and (\ref{theta_i}), respectively (cf.\ eqs.~(14) and (20) of
\cite{BDU}). The Clausen function is defined as
$\Cl{2}{\theta}=\mbox{Im}\left[\Li{2}{e^{{\rm i}\theta}}\right]$
(for details, see e.g.\ in \cite{Lewin}).
Note that the combination of Clausen functions
(\ref{Clausens}) is related to the volumes of asymptotic
tetrahedra in hyperbolic space of constant curvature
(see in \cite{geom}).
Worth noting is that at the threshold the result for the symmetric
combination  (\ref{S221}) is
\be
\label{S221T}
S^{(T)}_{221}=
-\frac{\pi^{4-2\ep}\;\Gamma^2(1+\ep)}
{(m_1m_2m_3)^{1/2+3\ep} \mu_T^{1/2-5\ep}}
\left[\frac{\pi}{2\ep}
+\pi {\cal{L}}(m_1,m_2,m_3)
+{\cal{C}}(\theta_1,\theta_2,\theta_3)
\right]
+ {\cal{O}}(\ep) .
\ee
Once more, we would like to emphasize that the $1/\ep$ poles 
in eqs.~(\ref{Lt221})
and (\ref{S221T}) (accompanied by an extra factor of $\pi$)
correspond to the threshold singularity: there are no
ultraviolet divergences in these results.

Using recurrence relations, we can obtain
results for $L_T(4-2\ep;1,1,3)$ with different masses,
as well as for the higher integrals.
Some relevant results are collected in Appendix~B.

In the equal-mass case,
$\theta_1=\theta_2=\theta_3={\textstyle{1\over3}}\pi$
(cf.\ eq.~(\ref{theta_i})), $T^{-}(1)=0$
(cf.\ eq.~(\ref{T-})),
${\cal{C}}(\pi/3,\pi/3,\pi/3)=5\Cl{2}{\pi/3}$
(we take into account that $\Cl{2}{2\pi/3}=
{\textstyle{2\over3}}\Cl{2}{\pi/3}$).
Therefore we get
\bea
\label{Lt221eq}
\left. L_T(n;2,2,1)\right|_{m_1=m_2=m_3\equiv m}
\equiv L_T^{\rm (eq)}(n;2,2,1)
\hspace{65mm}
\nn \\
= - \pi^{4-2\ep}\;
\Gamma^2(1+\ep) \; m^{-2-4\ep}
\left\{ \frac{\pi}{6\sqrt{3}} \left( \frac{1}{\ep}
-\ln{3} - 6\ln{2} \right)
+ \frac{5}{3\sqrt{3}} \Cl{2}{\frac{\pi}{3}} \right\}
+ {\cal{O}}(\ep).
\eea
It is interesting that $\Cl{2}{\pi/3}$ (i.e.\ the maximal
value of $\Cl{2}{\theta}$) also appears in the 
two-loop integrals containing a threshold 
singularity\footnote{For some other examples illustrating
occurence of the Clausen function in two-loop calculations,
including vacuum integrals and the pseudothreshold values,
see e.g.\ in \cite{Cl2,DT2}. Note that 
$\Cl{2}{\pi/3}=
\left[\psi'({\textstyle{1\over3}})-
{\textstyle{2\over3}}\pi^2\right]/(2\sqrt{3})$ 
(cf.\ in \cite{DT2,Can}).}.

In fact, knowing the result (\ref{Lt221eq}) 
we can obtain $\ep$-part of the integrals
$L_T^{\rm (eq)}(n;1,1,2)$ and $L_T^{\rm (eq)}(n;1,1,1)$.
Using recurrence relations, we get
\be
L_T^{\rm (eq)}(n;1,1,2)=
\frac{1}{(n\!-\!3)(3n\!-\!10)}
\left\{ -8 m^2 (n-4) L_T^{\rm (eq)}(n;2,2,1)
+ (2n-7) L^{\rm (eq)}(2,2,0)\right\} .
\ee
It is important that the integral (\ref{Lt221eq}) enters with a
factor $(n-4)$. Expanding in $\ep$ we get
\bea
L_T^{\rm (eq)}(4-2\ep;1,1,2)&=&
\pi^{4-2\ep}\; \Gamma^2(1+\ep)\; m^{-4\ep}
\bigg\{ -\frac{1}{2\ep^2}-\frac{1}{2\ep}+\frac{1}{2}+\frac{11}{2}\ep
\nn \\
&& -\frac{4\pi}{3\sqrt{3}}
\big[ 1+5\ep-\ep\ln{3}-6\ep\ln{2} \big]
-\frac{40\ep}{3\sqrt{3}}\Cl{2}{\frac{\pi}{3}} \bigg\}
+{\cal{O}}(\ep^2) .
\hspace{4mm}
\eea
Furthermore, using eq.~(\ref{Tar111}) we obtain
\be
L_T^{\rm (eq)}(n;1,1,1)=\frac{6 m^2}{(n-3)(3n-8)}
\left\{ 4(n-3)\; L_T^{\rm (eq)}(n;1,1,2) - L^{\rm (eq)}(n;2,2,0) \right\} .
\ee
Finally, we arrive at
\bea
L_T^{\rm (eq)}(4-2\ep;1,1,1)=
\pi^{4-2\ep}\; \Gamma^2(1+\ep)\; m^{2-4\ep}
\left\{ -\frac{3}{2\ep^2}-\frac{9}{4\ep}+\frac{45}{8}+\frac{855}{16}\ep
\right.
\hspace{20mm}
\nn \\
\left.
+\frac{4\pi}{\sqrt{3}}
\left[ -2-13\ep+2\ep\ln{3}+12\ep\ln{2} \right]
-\frac{80\ep}{\sqrt{3}}\Cl{2}{\frac{\pi}{3}} \right\}
+{\cal{O}}(\ep^2) .
\hspace{4mm}
\eea

\newpage

{\large\bf \noindent 5. Evaluation of the (p-p) contribution}

\vspace{3mm}

According to the prescription of \cite{BS},
we should start by expanding the propagators
in the integrand of eq.~(\ref{SS1}) in $p_{i0}^2$,
\be
\label{exp2}
\frac{1}{\left[p_{i0}^2-\vec{p}_i^2
         +2\xi_i (k p_i)-\xi_i^2 y\right]^{\nu_i}}
\Rightarrow
\sum_{j_i=0}^{\infty}\frac{(\nu_i)_{j_i}}{j_i!}\;
\frac{p_{i0}^{2j_i}}
{\left[-\vec{p}_i^2
         +2\xi_i k_0 p_{i0} -\xi_i^2 y\right]^{\nu_i+j_i}} \; ,
\ee
where, as before, we imply that $p_3=-p_1-p_2$.
However, it happens inconvenient to
expand integrand in the time components of the loop momenta
(according to eq.~(\ref{exp2}))
and then take $p_{10}$ and $p_{20}$ integrals, because
one arrives at rather cumbersome integrals in $\vec{p}_1$
and $\vec{p}_2$. We shall proceed in a slightly different way.

Let us instead exponentiate, by use of
the alpha parameters, every propagator of the unexpanded Feynman
integral (\ref{SS1}) and perform
integration in $\vec{p}_1$ and $\vec{p}_2$.
In particular, for the integral $L(4-2\ep;1,1,1)$ we have
\bea
-{\mbox{i}}^{2\ep} \pi^{n-1}
\int\limits_0^\infty
\int\limits_0^\infty
\int\limits_0^\infty
\frac{{\mbox{d}} \alpha_1 {\mbox{d}} \alpha_2 {\mbox{d}} \alpha_3}
{(\alpha_1 \alpha_2 +\alpha_2 \alpha_3+\alpha_3 \alpha_1)^{3/2-\ep}}
\exp\left[
-\mbox{i}y(\xi_1^2\alpha_1+\xi_2^2\alpha_2+\xi_3^2\alpha_3)\right]
\nn \\
\times \int\limits_{-\infty}^{\infty}
\int\limits_{-\infty}^{\infty}
{\mbox{d}} p_{10} {\mbox{d}} p_{20}
\exp\left\{\mbox{i}[(\alpha_1+\alpha_3)p_{10}^2
+2 \alpha_3 p_{10}p_{20}
+ (\alpha_2+\alpha_3)p_{20}^2]\right\}
\nn \\
\times \exp\left\{ \mbox{i}
\left[ 2 k_0 p_{10} (\xi_1 \alpha_1 - \xi_3 \alpha_3)
+ 2 k_0 p_{20} (\xi_2 \alpha_2 - \xi_3 \alpha_3) \right] \right\} \, .
\eea
Only now we perform Taylor expansion: namely, in the exponent with
$p_{10}^2$, $p_{20}^2$ and $p_{10}p_{20}$.
Then the evaluation of the integral of an arbitrary resulting term
is simple: we perform integrations in $p_{10}$ and $p_{20}$ and
obtain terms proportional to derivatives of two $\delta$-functions,
$\delta\big( 2 k_0 (\xi_1 \alpha_1 - \xi_3 \alpha_3)\big)$ and
$\delta\big( 2 k_0 (\xi_2 \alpha_2 - \xi_3 \alpha_3)\big)$.
Then, we take the integrals in $\alpha_1$ and $\alpha_2$, and finally
in $\alpha_3$. As a result we have obtained an explicit formula
for the (p-p) contribution in arbitrary order,
written through a finite eight-fold sum.

In eq.~(\ref{hhpp})
the (p-p) contribution starts from the order $y^2$, so that
$C_0^{{\rm (p-p)}}=C_1^{{\rm (p-p)}}=0$. For the first two non-trivial
orders we have
\bea
C_2^{{\rm (p-p)}} &=&
\frac{\pi \left( \xi_1 \xi_2 \xi_3 \right)^{1/2-\ep}}
     {4\ep(1-\ep)(1-2\ep)} \ ,
\label{ppLO}
\\
C_3^{{\rm (p-p)}} &=&
\frac{\pi \left( \xi_1 \xi_2 \xi_3 \right)^{-1/2-\ep}}
     {64 \ep(1-\ep)(1-2\ep)}
\left[ (1\!-\!2\ep) \left(\xi_1\xi_2\!+\!\xi_2\xi_3\!+\!\xi_3\xi_1\right)
-(5\!-\!18\ep) \xi_1 \xi_2 \xi_3 \right] .
\hspace{6mm}
\label{ppNLO}
\eea

In the equal-mass case, we get
\be
\label{ajbj}
C_j^{{\rm (p-p),(eq)}}=(-1)^{j+1} \frac{\pi}{\sqrt{3}}
\left[ a_j\left(\frac{1}{\ep}+3\ln{3}\right)
+b_j\right] + {\cal{O}}(\ep) \; .
\ee
The coefficients $a_j$ and $b_j$ ($j\leq12$) are given in Table~1.
Note that the threshold expansion of the imaginary part
of the sunset diagram is
completely characterized by the pole part of the (p-p) contribution,
i.e.\ by the coefficients $a_j$ in Table~1.
We have compared our results for the imaginary part with analytical
expressions (in terms of elliptic integrals)
presented in \cite{Bau} (cf.\ also in \cite{phase})
and found complete agreement.

\begin{table}[ht]
\caption{The coefficients in eqs.~(\ref{ajbj}), (\ref{ajbjcj}) 
         and (\ref{apjbpj}).}
\begin{center}
\begin{tabular}{|l||l|l||l|l|l|l|}   
\hline {\rule{0mm}{5mm}}
j & $a_j$ & $b_j$
& $\widetilde{a}_j=\widetilde{a}'_j$ & $\widetilde{b}_j$
& $\widetilde{b}'_j$ & $\widetilde{c}_j$ \\[2mm]
\hline \hline {\rule{0mm}{5mm}}
2 &
$\textstyle{1\over12}$ &
$\textstyle{1\over4}$ &
$\textstyle{1\over12}$ &
$\textstyle{2\over9}$ &
$\textstyle{1\over4}$ & 
-$\textstyle{1\over12}$ \\[2mm]
\hline {\rule{0mm}{5mm}}
3 &
$\textstyle{1\over48}$ &
$\textstyle{1\over16}$ &
$\textstyle{1\over16}$ &
$\textstyle{41\over144}$ &
$\textstyle{3\over16}$ &
$\textstyle{1\over8}$ \\[2mm]
\hline {\rule{0mm}{5mm}}
4 &
$\textstyle{7\over768}$ &
$\textstyle{15\over512}$ &
$\textstyle{13\over256}$ &
$\textstyle{1279\over4608}$ &
$\textstyle{79\over512}$ &
$\textstyle{11\over64}$ \\[2mm]
\hline {\rule{0mm}{5mm}}
5 &
$\textstyle{1\over192}$ &
$\textstyle{9\over512}$ &
$\textstyle{11\over256}$ &
$\textstyle{2407\over9216}$ &
$\textstyle{17\over128}$ &
$\textstyle{59\over320}$ \\[2mm]
\hline {\rule{0mm}{5mm}}
6 &
$\textstyle{83\over24576}$ &
$\textstyle{583\over49152}$ &
$\textstyle{305\over8192}$ &
$\textstyle{72019\over294912}$ &
$\textstyle{5767\over49152}$ &
$\textstyle{1901\over10240}$ \\[2mm]
\hline {\rule{0mm}{5mm}}
7 &
$\textstyle{233\over98304}$ &
$\textstyle{1693\over196608}$ &
$\textstyle{1073\over32768}$ &
$\textstyle{269359\over1179648}$ &
$\textstyle{20719\over196608}$ &
$\textstyle{261607\over1433600}$ \\[2mm]
\hline {\rule{0mm}{5mm}}
8 &
$\textstyle{1381\over786432}$ &  
$\textstyle{6889\over1048576}$ &
$\textstyle{7623\over262144}$ &
$\textstyle{1345217\over6291456}$ &
$\textstyle{100361\over1048576}$ &
$\textstyle{2034657\over11468800}$ \\[2mm]
\hline {\rule{0mm}{5mm}}
9 &
$\textstyle{2129\over1572864}$ &
$\textstyle{32695\over6291456}$ &
$\textstyle{13623\over524288}$ &
$\textstyle{52988741\over264241152}$ &
$\textstyle{183713\over2097152}$ &
$\textstyle{27534999\over160563200}$ \\[2mm]
\hline {\rule{0mm}{5mm}}
10 &
$\textstyle{108257\over100663296}$ &
$\textstyle{1701199\over402653184}$ &
$\textstyle{781899\over33554432}$ &
$\textstyle{3184520135\over16911433728}$ &
$\textstyle{10819013\over134217728}$ &
$\textstyle{1697828907\over10276044800}$ \\[2mm]
\hline {\rule{0mm}{5mm}}
11 &
$\textstyle{352373\over402653184}$ &
$\textstyle{5653147\over1610612736}$ &
$\textstyle{2809445\over134217728}$ &
$\textstyle{35912515043\over202937204736}$ &
$\textstyle{119896193\over1610612736}$ &
$\textstyle{14363946379\over90429194240}$ \\[2mm]
\hline {\rule{0mm}{5mm}}
12 &
$\textstyle{4677635\over6442450944}$ &
$\textstyle{382328867\over128849018880}$ &
$\textstyle{40375137\over2147483648}$ &
$\textstyle{900245481797\over5411658792960}$ &
$\textstyle{2959291009\over42949672960}$ &
$\textstyle{220597949391\over1446867107840}$ \\[2mm]
\hline
\end{tabular}
\end{center}
\end{table}

Note that for an arbitrary,
$L$-loop ``water-melon'' diagram (consisting of two
vertices and $(L+1)$ massive lines between them)
the (p-p-\ldots-p) contribution
can also be analytically calculated in every order.
In particular, in the leading order we have the following result:
\be
{\mbox{i}}^L\pi^{(n+1)L/2}
\Gamma\big((\ep-3/2)L+1\big)
\left(\prod_{j=1}^{L+1} \xi_j^{1/2-\ep}\right)
\frac{1}{(k^2)^{L/2} y^{(\ep-3/2)L+1}} \, ,
\label{gen2res}
\ee
where, by analogy with eq.~(\ref{xi_i}),
$\xi_j=m_j/(\sum_{i=1}^{L+1}m_i)$.

\vspace{5mm}

{\large\bf \noindent 6. Results and numerical comparison}

\vspace{3mm}

Collecting parts of the algorithm described in the previous
sections, we obtain terms of the threshold expansion
of the sunset diagram (\ref{def_L}).
To see how efficiently the threshold expansion works,
we have compared our results with a semi-numerical program
based on the algorithm of \cite{PT}\footnote{We are 
grateful to P.~Post and J.B.~Tausk for kind permission
to use their {\sf REDUCE} program for the comparison.}.
The ultraviolet-divergent terms were compared analytically,
whereas the finite (in $\ep$) part was treated numerically,
using an integral representation from \cite{PT}.

Our approximations correspond to the threshold expansion
given by eqs.~(\ref{hhpp}) and (\ref{hhpp2}). 
We refer to the sum of terms up to order $y^N$ as to 
an $N$-th approximation. We illustrate the convergence
in the plots, Fig.~2 and Fig.~3, where we present the subtracted
(i.e., without poles in $\ep$) real part and the imaginary part  
of $(\mu_T^2)^{-1+2\ep}L_T(4-2\ep;1,1,1)$. 
Note that the imaginary part is finite as $\ep\to0$.
These quantities are shown as functions of $k^2/\mu_T^2$
(which equals $k^2/(9m^2)$ in the equal-mass case).
In all plots, the highest-order curve is denoted 
by the solid line, whereas the lower approximations are
drawn by various dashed lines (see the plots).
The results of the above-mentioned numerical
program \cite{PT} are displayed as crosses.

As the first example for numerical comparison, we have chosen 
the case $\xi_1=0.1$, $\xi_2=0.3$, $\xi_3=0.6$, when the masses
are essentially different (but still,
one cannot neglect either of them).
Here, the expansion (\ref{hhpp2}) in terms of $y/\mu_T^2$ is used.
The results are shown in Fig.~2. On one hand,
we can see that the expansion of the real part converges 
reasonably well. On the other hand, 
the expansion of the imaginary part breaks down just  
above $k^2/\mu_T^2\simeq1.5$. In fact, 
the same happens to the real part, when $k^2/\mu_T^2\simeq2.2$.
This is because the expansion
parameter $y/\mu_T^2$ is already not so small. 

If we use, for $k^2>\mu_T^2$, the expansion in terms of $y/k^2$ 
the convergence is much better. 
For instance, at $k^2=2\mu_T^2$ the $N=7$ approximation
provides a four-digit precision for the imaginary part.  
In general, it appears that an optimal
choice is to use the expansion in $y/\mu_T^2$ (cf.\ eq.~(\ref{hhpp2}))
for $k^2<\mu_T^2$, and the expansion in $y/k^2$
(cf.\ eq.~(\ref{hhpp})) for $k^2>\mu_T^2$. 
For the imaginary part, it is enough to use only the $y/k^2$ 
expansion, since it vanishes for $k^2<\mu_T^2$.

As the second example, let us
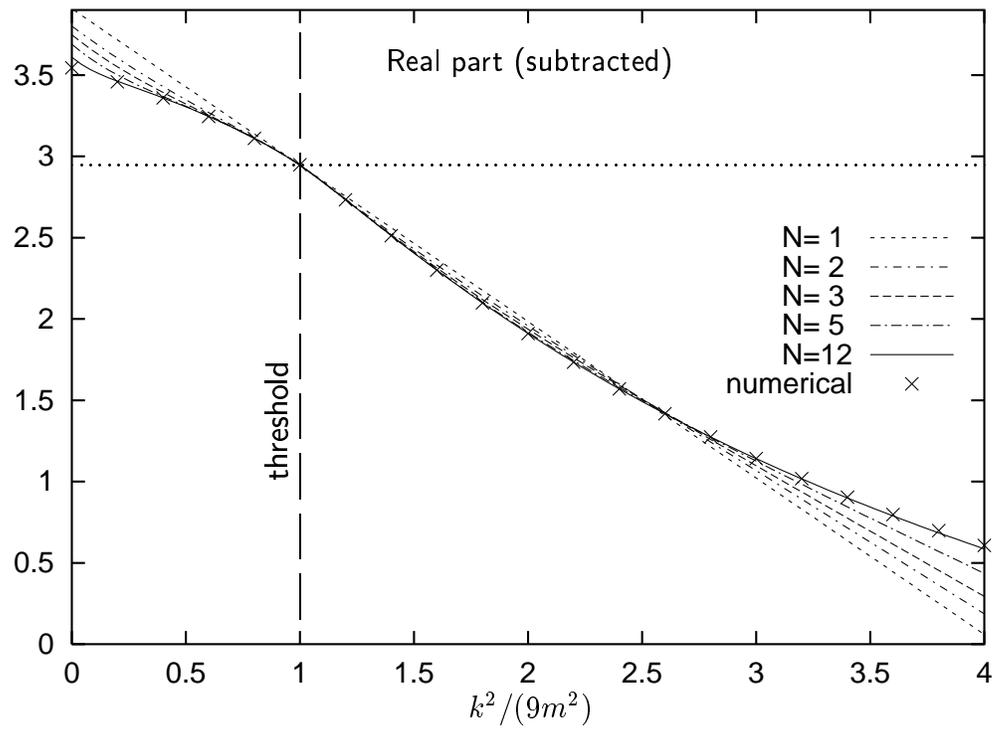
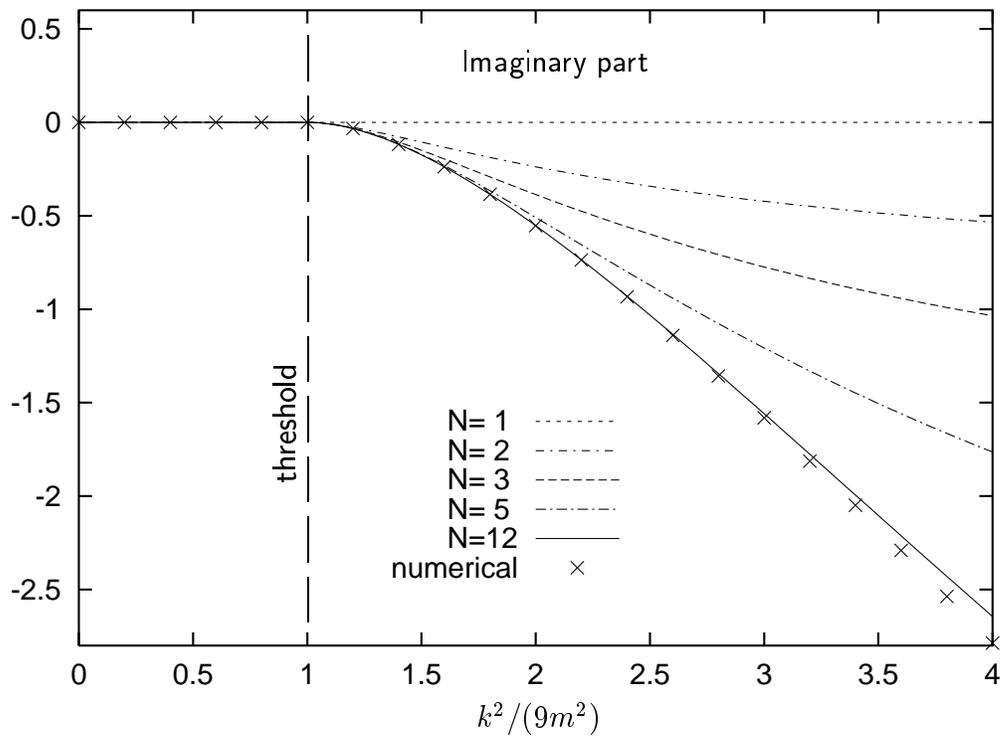
consider eq.~(\ref{hhpp2}) in the case of equal masses,
$m_1=m_2=m_3\equiv m$ ($\xi_i=1/3$), remembering that
$\widetilde{C}_0^{{\rm (p-p)}}=\widetilde{C}_1^{{\rm (p-p)}}=0$.
The two lowest (h-h) contributions are
\bea
\widetilde{C}_0^{{\rm (h-h)},{\rm (eq)}}
&=& 3^{4\ep}\left[ \frac{1}{6\ep^2}+\frac{1}{4\ep}-\frac{5}{8}
                 +\frac{8\pi}{9\sqrt{3}} \right]
+ {\cal{O}}(\ep),
\\
\widetilde{C}_1^{{\rm (h-h)},{\rm (eq)}}
&=& 3^{4\ep}\left[ \frac{1}{4\ep}+\frac{23}{24}
                -\frac{4\pi}{9\sqrt{3}} \right]
+ {\cal{O}}(\ep).
\eea
It is easy to check that these two terms absorb all ultraviolet
singularities of the original sunset integral.
For $j\geq 2$, the coefficients have the following form:
\bea
\label{ajbjcj}
\widetilde{C}_j^{{\rm (h-h)},{\rm (eq)}}
&=& \frac{\pi}{\sqrt{3}}
\left[ \widetilde{a}_j
\left(\frac{1}{\ep} + 3\ln 3 - 6\ln 2\right)
+ \widetilde{b}_j \right]
+ \frac{10\widetilde{a}_j}{\sqrt{3}} \Cl{2}{\frac{\pi}{3}}
- \widetilde{c}_j + {\cal{O}}(\ep) ,
\\
\label{apjbpj}
\widetilde{C}_j^{{\rm (p-p)},{\rm (eq)}}
&=& - \frac{\pi}{\sqrt{3}}\left[
\widetilde{a}'_j \left( \frac{1}{\ep} + 3\ln 3 \right)
+ \widetilde{b}'_j \right] + {\cal{O}}(\ep) .
\eea

The $1/\ep$ poles in the (h-h) contribution, eq.~(\ref{ajbjcj}),  
correspond to the threshold singularity.
However, since there is no threshold singularity in the
case $\nu_1=\nu_2=\nu_3=1$, these poles should cancel
with those from the (p-p) contribution, eq.~(\ref{apjbpj}).
This is the case if $\widetilde{a}'_j=\widetilde{a}_j$.
We have checked that this property is valid for all available
contributions. The corresponding coefficients
(for $j\leq 12$) are collected in Table~1.

As a result, using eq.~(\ref{hhpp2}) we get the following threshold
expansion in the equal-mass case:
\bea
\label{eq_mass_exp}
L^{{\rm (eq)}}(4-2\ep;1,1,1)=-\pi^{4-2\ep} \Gamma^2(1+\ep)
(9m^2)^{1-2\ep}
\bigg\{ \widetilde{C}_0^{{\rm (h-h),(eq)}}
+\frac{y}{9m^2} \widetilde{C}_1^{{\rm (h-h),(eq)}}
\hspace{17mm}
\nn \\
+ \frac{2}{\sqrt{3}} 
\bigg[ \pi \ln\frac{y}{9m^2} -\! 3\pi\ln{2} 
      \!+\! 5 \Cl{2}{\frac{\pi}{3}} \bigg]
A\left(\frac{y}{9m^2}\right)
\!+\frac{\pi}{\sqrt{3}} B\left(\frac{y}{9m^2}\right)
\!-C\left(\frac{y}{9m^2}\right)\bigg\}
\!+\!{\cal{O}}(\ep),
\hspace{4mm}
\eea
with the ``form factors''
\be
\label{ff}
A(z)\equiv \sum_{j=2}^{\infty} \widetilde{a}_j z^j,
\hspace{10mm}
B(z)\equiv \sum_{j=2}^{\infty} 
(\widetilde{b}_j-\widetilde{b}'_j) z^j,
\hspace{10mm}
C(z)\equiv \sum_{j=2}^{\infty} \widetilde{c}_j z^j .
\ee
Note that the imaginary part of the expression in braces
in (\ref{eq_mass_exp}) equals
\be
\frac{2\pi^2}{\sqrt{3}} A\left(\frac{y}{9m^2}\right)
\theta(k^2-9m^2) ,
\ee
where the function $A(z)$ can be expressed in terms 
of elliptic integrals \cite{Bau}. 

We have also compared this equal-mass example with 
the program based on the algorithm of \cite{PT}.
The results of this numerical comparison are shown in Fig.~3.
We have used the ``combined'' expansion: 
below the threshold (for $k^2<\mu_T^2$) 
our approximations correspond to the
expansion (\ref{eq_mass_exp})--(\ref{ff}), whereas
beyond the threshold (for $k^2>\mu_T^2$) we switch to the
expansion in $y/k^2$. Technically, this transformation can be done
just by substituting the arguments of the functions (\ref{ff})
via $y/(\mu_T^2)\to(y/k^2)/(1+y/k^2)$ and re-expanding
in $y/k^2$, up to the given order $N$.

{} From the plots it is clear that subsequent approximations 
are getting better in a wide range of the values of the external
momentum squared, i.e.\ the threshold expansion indeed works well.
Close to the threshold, it is enough to take just
a few terms of the expansion for achieving 
precision which is better than that of the numerical
program \cite{PT}. 
For $k^2=0.5\mu_T^2$, the $N=7$ and $N=12$ approximations to the
real part reproduce four and six digits, respectively.
When $k^2=2\mu_T^2$, the $N=6$ approximation gives us three 
digits, whereas the $N=12$ curve provides six-digit precision.

Note that the threshold expansion (at least, in its present form)
does not seem to work in Euclidean region, i.e.\ when $k^2<0$. 
Then, the limits when some of the particles become massless
are also non-regular, since one should take into account
ultrasoft regions (see in Section~1). However, for these
cases (when some masses vanish) exact results are available 
in refs.~\cite{B+ST,B+T}. 

\newpage

{\large\bf \noindent 7. Conclusions} 

\vspace{3mm}

Let us briefly summarize the main results of this paper.

We have considered the application of prescriptions of 
ref.~\cite{BS} to the construction of the threshold expansion
of the sunset diagram (Fig.~1) at the three-particle
threshold. We have treated the most complicated case, when 
all particles involved in the threshold cut are massive.
The expansion involves two contributions: the (h-h)
and the (p-p) ones.

The main technical difficulties are related to the 
(h-h) contribution. Basically, it corresponds 
to a formal Taylor expansion around the threshold.
To construct the coefficients of this expansion, one needs
to calculate threshold values of the corresponding sunset
integrals (\ref{def_L}) with higher powers of the propagators.
Although results for the lowest integrals were known
up to the finite parts in $\ep$ \cite{BDU}, the recursive 
procedure (based on the techniques \cite{ibp,Tarasov})
required either $\ep$-part of those integrals or an
explicit result for one of the higher integrals
(e.g.\ $L_T(4-2\ep;2,2,1)$). 
We have calculated this integral analytically, in terms of
dilogarithms and Clausen functions (\ref{Lt221}).
Note that it contains a $1/\ep$ pole which is associated
with the threshold singularity (treated in the framework
of dimensional regularization).
The result (\ref{Lt221}) has completed the algorithm for 
calculating higher terms of the (h-h) contribution.

Collecting the (h-h) and (p-p) contributions, we get 
the terms of the threshold expansion.
A comparison with the results of semi-numerical program \cite{PT}
has been performed. It was shown that the combined expansion
(using the variables $y/\mu_T^2$ and $y/k^2$ below and above
the threshold, respectively) provides good analytical 
approximations in a wide region of values of the 
external momentum squared.
This illustrates how efficiently the general procedure of threshold
expansion \cite{BS} works for diagrams with three-particle  
thresholds in the totally massive case.

\vspace{4mm}

\noindent {\bf Acknowledgments.}
We are grateful to F.A.~Berends for useful discussions and kind
hospitality at the Instituut-Lorentz, University of Leiden where
an essential part of this work has been done, during our visits
supported by the EU INTAS grant 93--0744. We would like 
to thank O.V.~Tarasov and J.B.~Tausk  for useful comments.
A.~D.'s research has been essentially supported by the Alexander
von Humboldt
Foundation, whereas the work of V.~S. has been supported
by the Volkswagen Foundation under contract No.~I/73611.
Both of us acknowledge partial support from
the Russian Foundation for Basic Research, project 98--02--16981.

\vspace{5mm}

{\large\bf Appendix A: Recurrence relations for different masses}

\vspace{3mm}

The main parameter of recursion is the sum of the indices,
$\sigma\equiv\nu_1+\nu_2+\nu_3$. Under ``lower integrals''
we understand the integrals with lower values of $\sigma$.
When one of the $\nu$'s is zero, the corresponding integral
is trivial (see eq.~(\ref{tadpoles})).
The integrals with $\sigma=3$ ($\nu_1=\nu_2=\nu_3=1$)
and $\sigma=4$ ($\nu_1=\nu_2=1$, $\nu_3=2$ and permutations)
are known \cite{BDU}. We are therefore interested in
calculation of the integrals with $\sigma\geq 5$.

Using eqs.~(\ref{++13}), (\ref{++23}) we can express any
integral $L_T(n;\nu_1,\nu_2,\nu_3)$ with positive $\nu$'s in terms of
$L_T(n;1,1,\sigma-2)$, $L_T(n;2,1,\sigma-3)$,
$L_T(n;1,2,\sigma-3)$ and $L_T(n;2,2,\sigma-4)$.
Furthermore, using the integration-by-parts identities
we can express $L_T(n;2,1,\sigma-3)$ and $L_T(n;1,2,\sigma-3)$
in terms of $L_T(n;1,1,\sigma-2)$, $L_T(n;2,2,\sigma-4)$
and lower integrals, via
\bea
&& \hspace*{-20mm}
2 (\nu_3-1) m_3 (m_1+m_2) (2 m_1 m_2+m_1 m_3+m_2 m_3)
L_T(n;2,1,\nu_3)
\nn \\
&=& 2 m_2^2 (m_1+m_2) (m_1+m_3) L_T(n;2,2,\nu_3-1)
\nn \\
&& +2 m_3^2 \nu_3 (\nu_3-1) (m_1+m_2) (m_2+m_3) L_T(n;1,1,\nu_3+1)
\nn \\
&& -(\nu_3-1) (2 n-2\nu_3-3) (m_1+m_2) (m_2+m_3) L_T(n;1,1,\nu_3)
\nn \\
&& -(n-\nu_3-1) (m_1 m_2+m_1 m_3+m_2 m_3) L_T(n;2,1,\nu_3-1)
\nn \\
&& -(n-\nu_3-1) m_2^2 L_T(n;1,2,\nu_3-1)
 +(\nu_3-1) m_2^2 L(n;0,2,\nu_3)
\nn \\
&& +(\nu_3-1) (m_1 m_2+m_1 m_3+m_2 m_3) L(n;2,0,\nu_3) ,
\eea
and an analogous equation for $L_T(n;1,2,\nu_3)$
(only $m_1$ and $m_2$, together with the arguments of the integrals
corresponding to $\nu_1$ and $\nu_2$, are to be permuted).

Then, considering equations (\ref{Tar}) (which follow from
ref.~\cite{Tarasov}), we obtain the following
solution for the remaining integrals (with $\nu_1=\nu_2=1$
and $\nu_1=\nu_2=2$):
\bea
\label{Rec11}
&& \hspace*{-8mm}
32 m_3^2 (\nu_3\!-\!1) (\nu_3\!-\!2) (n\!-\!\nu_3\!-\!1)
(m_1\!+\!m_2) (m_2\!+\!m_3) (m_3\!+\!m_1) \mu_T
L_T(n;1,1,\nu_3)
\nn \\
&=&2 (\nu_3\!-\!2) (m_1\!+\!m_2)
 \bigg[ (3n\!-\!4\nu_3\!-\!1) (3n\!-\!4\nu_3\!+\!1)
 (m_1\!+\!m_3) (m_2\!+\!m_3) \mu_T
\nn \\
&&   +2 (n-3) (n-\nu_3-1) m_3 \mu_T (\mu_T+m_3)
\nn \\
&&
   +(n-3) (n-2\nu_3+1) m_3 (m_1+m_3) (m_2+m_3)
\bigg] L_T(n;1,1,\nu_3-1)
\nn \\
&& -2 (m_1+m_2)
\bigg[ (n-2\nu_3+1) (m_1+m_2) (m_2+m_3) (m_3+m_1)
\nn \\
&&
     + 2 (n-\nu_3-1) m_1 m_2 (\mu_T+m_3)
\bigg] L_T(n;2,2,\nu_3-3)
\nn \\
&& +(m_1+m_2) \bigg\{ (n-3) (m_1+m_3) (m_2+m_3)
\left[(2 n-11) (m_1+m_2)-2 m_3 \right]
\nn \\
&& +(n-\nu_3-1) (n+2\nu_3-9) m_1 m_2 m_3
\nn \\
&& +(n-\nu_3-1) \mu_T
\bigg[ (7 n-20 \nu_3+39) (m_1+m_3) (m_2+m_3)
\nn \\
&&
 -m_3 \bigg( 2 (n-3\nu_3+6) (m_1+m_2)+(3 n-14\nu_3+33) m_3 
\bigg) \bigg] \bigg\}
\nn \\
&& \times
\left[ L_T(n;2,1,\nu_3-2)+L_T(n;1,2,\nu_3-2)\right]
\nn \\
&& +(n-\nu_3) (m_1-m_2)
\bigg\{ -(n-3) (m_1+m_2) (m_1+m_3) (m_2+m_3)
\nn \\
&&
 +2 (n-\nu_3-1)
\left[ \mu_T
(2 m_1 m_2+3 m_1 m_3+3 m_2 m_3+m_3^2)
-m_1 m_2 m_3 \right]
\bigg\}
\nn \\
&& \times
\left[ L_T(n;2,1,\nu_3-2)-L_T(n;1,2,\nu_3-2) \right]
\nn \\
&& -2 (\nu_3-2) (m_1+m_3)
\bigg\{ (m_1+m_2) (m_2+m_3) \left[ m_1 (n-5)-m_3 (n-3) \right]
\nn \\
&&
 +2 m_3 (n-\nu_3-1)
\big[ \mu_T (m_1+m_2)  
+m_1 (m_2+m_3) \big]
\bigg\}
L(n;2,0,\nu_3-1)
\nn \\
&& -2 (\nu_3-2) (m_2+m_3)
\bigg\{ (m_1+m_2) (m_1+m_3) \left[ m_2 (n-5)-m_3 (n-3) \right]
\nn \\
&&
+2 m_3 (n\!-\!\nu_3\!-\!1)
\big[ (m_1+m_2) \mu_T
+m_2 (m_1+m_3) \big]
\bigg\}
L(n;0,2,\nu_3-1) ,
\eea
\bea
\label{Rec22}
&& \hspace*{-8mm}
32 (n-\nu_3-3) m_1 m_2 (m_1+m_2) (m_2+m_3) (m_3+m_1) \mu_T
L_T(n;2,2,\nu_3)
\nn \\
&=& 2(n-3)\nu_3 (m_1+m_2)
\bigg\{ (m_1+m_3) (m_2+m_3)
\left[ (n-5) (m_1+m_2)-2 m_3\right]
\nn \\
&&
-2 m_3 (n-\nu_3-3) \left[ \mu_T^2
-m_1 m_2\right]\bigg\}
L_T(n;1,1,\nu_3+1)
\nn \\
&& +2 (m_1+m_2) \bigg[ (n-3) (m_1+m_2) (m_2+m_3) (m_3+m_1)
\nn \\
&&
+2 (n-\nu_3-3) \mu_T
(2 m_1 m_2+m_1 m_3+m_2 m_3)
\bigg] L_T(n;2,2,\nu_3-1)
\nn \\
&& +(m_1+m_2)
\bigg\{ (n-3) (m_1+m_3) (m_2+m_3)
\left[ (2 n-11) (m_1+m_2)-2 m_3 \right]
\nn \\
&& +(n-\nu_3-3) (n+2\nu_3-5) m_1 m_2 m_3
\nn \\
&& +(n-\nu_3-3) \mu_T
\bigg[ -3 (3 n-4\nu_3-5) (m_1+m_3) (m_2+m_3)
\nn \\
&&
+m_3 \bigg( 2 (n-3\nu_3) (m_1+m_2)+5 (n-2\nu_3-1) m_3 \bigg)
\bigg] \bigg\}
\nn \\
&& \times\left[ L_T(n;2,1,\nu_3)+L_T(n;1,2,\nu_3) \right]
\nn \\
&& +(n-\nu_3-2) (m_1-m_2)
\bigg\{ -(n-3) (m_1+m_2) (m_1+m_3) (m_2+m_3)
\nn \\
&&
+2 (n-\nu_3-3) \left[ -m_1 m_2 m_3
+ \mu_T
(2 m_1 m_2+m_3\mu_T)
\right]
\bigg\}
\nn \\
&& \times\left[ L_T(n;2,1,\nu_3)-L_T(n;1,2,\nu_3) \right]
\nn \\
&& +2\nu_3 (m_1+m_3)
\bigg\{ (m_1+m_2) (m_2+m_3)
\left[(n-3) m_3-(n-5) m_1\right]
\nn \\
&&
+2 m_3 (n-\nu_3-3) (m_1^2+m_2\mu_T)
\bigg\}
L(n;2,0,\nu_3+1)
\nn \\
&& +2\nu_3 (m_2+m_3)
\bigg\{ (m_1+m_2) (m_1+m_3)
\left[(n-3) m_3-(n-5) m_2\right]
\nn \\
&& +2 m_3 (n-\nu_3-3) (m_2^2+m_1\mu_T)
\bigg\}
L(n;0,2,\nu_3+1) .
\eea

We note that the factors $(n-\nu_3-1)$ (on the l.h.s.\ of
eq.~(\ref{Rec11})) and
$(n-\nu_3-3)$ (on the l.h.s.\ of eq.~(\ref{Rec22}))
yield $(n-4)$ at $\nu_3=3$ and $\nu_3=1$, respectively.
This illustrates the problem arising when getting the results
for $L_T(n;2,2,1)$ and $L_T(n;1,1,3)$ (see in section~3).

\vspace{5mm}

{\large\bf Appendix B: Threshold results for some other integrals}

\vspace{3mm}

Using eqs.~(\ref{L113_rec}), (\ref{s1}) and (\ref{S221T}), we obtain
\bea
L_T(4-2\ep;1,1,3)
=\frac{\pi^{4-2\ep}\;\Gamma^2(1+\ep)}{m_3^2}
\bigg\{ m_3^{-4\ep}
\left[\frac{1}{2\ep}+1-\frac{1}{\mu_T}
\left(m_1\ln\frac{m_1}{m_3}+m_2\ln\frac{m_2}{m_3}\right)
\right]
\nn \\
-\frac{(m_1m_2m_3)^{1/2-3\ep}}{2\mu_T^{3/2-5\ep}}
\left[\frac{\pi}{2\ep}+4\pi
+\pi {\cal{L}} + {\cal{C}}
\right] \bigg\}
+ {\cal{O}}(\ep) .
\hspace{6mm}
\eea
Here and below, ${\cal{L}}\equiv{\cal{L}}(m_1,m_2,m_3)$ and
${\cal{C}}\equiv{\cal{C}}(\theta_1,\theta_2,\theta_3)$,
see eqs.(\ref{Logs}) and (\ref{Clausens}).

Employing explicit recurrence relations from
Appendix~A, we can obtain results for the integrals $L_T$
with higher values of $\sigma=\nu_1+\nu_2+\nu_3$.
For instance, for $\sigma=6$ we get
\bea
\label{Lt222}
L_T(4-2\ep;2,2,2)=\frac{1}{4}\; \pi^{4-2\ep}\; \Gamma^2(1+\ep)\;
(m_1 m_2 m_3)^{-1-3\ep}\;
\mu_T^{-1+5\ep}
\hspace{44.5mm}
\nn \\
\times\!\left\{
-\frac{(m_1+m_2)(m_2+m_3)(m_3+m_1)+4m_1m_2m_3}
      {4\sqrt{m_1m_2m_3\mu_T^3}}
\left[\frac{\pi}{2\ep}+3\pi
+\pi {\cal{L}} + {\cal{C}}
\right]
\right.
\hspace{24mm}
\nn \\
+\pi\sqrt{\frac{m_1m_2m_3}{\mu_T^3}}
\left[ 3+\frac{m_1m_2m_3+\mu_T^3}{(m_1+m_2)(m_2+m_3)(m_3+m_1)}
\right] +1
\hspace{45mm}
\nn \\
+\frac{m_3(m_1-m_2)}{(m_2\!+\!m_3)(m_3\!+\!m_1)}\ln\frac{m_1}{m_2}\!
+\frac{m_1(m_2-m_3)}{(m_3\!+\!m_1)(m_1\!+\!m_2)}\ln\frac{m_2}{m_3}\!
+\frac{m_2(m_3-m_1)}{(m_1\!+\!m_2)(m_2\!+\!m_3)}\ln\frac{m_3}{m_1}
\nn \\
\left.
-\frac{1}{2\mu_T}
\left[ (m_1\!-\!m_2)\ln\frac{m_1}{m_2}+(m_2\!-\!m_3)\ln\frac{m_2}{m_3}
+(m_3\!-\!m_1)\ln\frac{m_3}{m_1} \right]
\right\}
+ {\cal{O}}(\ep),
\hspace{20mm}
\eea
\bea
\label{Lt123}
L_T(4-2\ep;1,2,3)=\frac{1}{8 m_2 m_3^2}\;\pi^{4-2\ep}\; \Gamma^2(1+\ep)\;
(m_1 m_2 m_3)^{-3\ep} \mu_T^{-1+5\ep}
\hspace{39mm}
\nn \\
\times\!\left\{
\frac{3(m_1+m_2)(m_2+m_3)(m_3+m_1)-4 m_1 m_3 \mu_T}
      {4\sqrt{m_1m_2m_3\mu_T^3}}
\left[\frac{\pi}{2\ep}+\frac{\pi}{3}
+\pi {\cal{L}} + {\cal{C}}
\right]
\right.
\hspace{10mm}
\nn \\
+\frac{\pi\; m_1}{\sqrt{m_1m_2m_3\mu_T}}
\left[\frac{2}{3}(3m_2\!-\!m_3)
-m_2 m_3
\frac{(m_1\!+\!m_2)(m_1\!+\!m_3)\!-\!m_2^2\!+\!m_3^2}
     {(m_1\!+\!m_2)(m_2\!+\!m_3)(m_3\!+\!m_1)}\right]
\nn \\
+ 1 + \frac{m_1(m_2-m_3)}{(m_1+m_2)(m_3+m_1)}\ln\frac{m_1}{m_2}
    + \frac{m_3(m_2-m_1)}{(m_2+m_3)(m_3+m_1)}\ln\frac{m_3}{m_2}
\nn \\
\left.
+\frac{3}{2\mu_T}
\left[(m_2+m_3)\ln\frac{m_1}{m_2}+(m_1+m_2)\ln\frac{m_3}{m_2}\right]
\right\}
+ {\cal{O}}(\ep).
\hspace{5mm}
\eea
The result for $L_T(4-2\ep;2,2,2)$, eq.~(\ref{Lt222}),
is totally symmetric in $m_1, m_2, m_3$, as it should.
Using eq.~(\ref{Lt123}) (and its permutations)
and the results for lower integrals, one can easily obtain result
for $L_T(4-2\ep;1,1,4)$ (and its permutations)
via eqs.~(\ref{++13})--(\ref{++23}).


\end{document}